\pdfoutput=1

\documentclass[12pt,a4paper]{article}

\usepackage{ifthen} 
\newboolean{pdflatex}
\setboolean{pdflatex}{true} 
\usepackage{graphicx}
\usepackage{epstopdf}
\usepackage{multirow}
\usepackage{rotating}
\usepackage{placeins}

\newboolean{articletitles}
\setboolean{articletitles}{true} 

\newboolean{uprightparticles}
\setboolean{uprightparticles}{false} 

\def\paperauthors{LHCb collaboration} 
\def\paperasciititle{Constraints on the K0->mumu branching fraction} 
\def\papertitle{Constraints on the $K^0_{\mathrm{S}}\rightarrow\mu^{+}\mu^{-}$ branching fraction} 
\def\paperkeywords{{High Energy Physics}, {LHCb}} 
\def\papercopyright{\the\year\ CERN for the benefit of the LHCb collaboration} 
\def\paperlicence{CC-BY-4.0 licence}
\def\paperlicenceurl{https://creativecommons.org/licenses/by/4.0/}

\usepackage[top=1in, bottom=1.25in, left=1in, right=1in]{geometry}

\usepackage{microtype}
\usepackage{lineno}  
\usepackage{xspace} 
\usepackage{caption} 

\usepackage{graphicx}  
\usepackage{color}
\usepackage{colortbl}
\graphicspath{{./figs/}} 
\DeclareGraphicsExtensions{.pdf,.PDF,png,.PNG}

\usepackage{amsmath} 
\usepackage{amssymb}
\usepackage{amsfonts}
\usepackage{upgreek} 

\newcommand*\patchAmsMathEnvironmentForLineno[1]{%
\expandafter\let\csname old#1\expandafter\endcsname\csname #1\endcsname
\expandafter\let\csname oldend#1\expandafter\endcsname\csname
end#1\endcsname
 \renewenvironment{#1}%
   {\linenomath\csname old#1\endcsname}%
   {\csname oldend#1\endcsname\endlinenomath}%
}
\newcommand*\patchBothAmsMathEnvironmentsForLineno[1]{%
  \patchAmsMathEnvironmentForLineno{#1}%
  \patchAmsMathEnvironmentForLineno{#1*}%
}
\AtBeginDocument{%
\patchBothAmsMathEnvironmentsForLineno{equation}%
\patchBothAmsMathEnvironmentsForLineno{align}%
\patchBothAmsMathEnvironmentsForLineno{flalign}%
\patchBothAmsMathEnvironmentsForLineno{alignat}%
\patchBothAmsMathEnvironmentsForLineno{gather}%
\patchBothAmsMathEnvironmentsForLineno{multline}%
\patchBothAmsMathEnvironmentsForLineno{eqnarray}%
}

\usepackage{hyperxmp}

\usepackage[pdftex,
            pdfauthor={\paperauthors},
            pdftitle={\paperasciititle},
            pdfkeywords={\paperkeywords},
            pdfcopyright={Copyright (C) \papercopyright},
            pdflicenseurl={\paperlicenceurl}]{hyperref}

\usepackage[colorinlistoftodos,textsize=scriptsize]{todonotes}

\usepackage[all]{hypcap} 

\usepackage{xspace} 
\usepackage{upgreek}

\def\lhcb   {\mbox{LHCb}\xspace}

\def\kloeii {\mbox{KLOE-2}\xspace}

\def\velo   {VELO\xspace}

\def\MagUp {\mbox{\em Mag\kern -0.05em Up}\xspace}

\ifthenelse{\boolean{uprightparticles}}%
{

 \def\Pmu         {\ensuremath{\upmu}\xspace}

 \def\Ppi         {\ensuremath{\uppi}\xspace}

 \def\PDelta      {\ensuremath{\Delta}\xspace}                 
 \def\PXi         {\ensuremath{\Xi}\xspace}                 
 \def\PLambda     {\ensuremath{\Lambda}\xspace}                 
 \def\PSigma      {\ensuremath{\Sigma}\xspace}                 
 \def\POmega      {\ensuremath{\Omega}\xspace}                 
 \def\PUpsilon    {\ensuremath{\Upsilon}\xspace}

 \def\PB      {\ensuremath{\mathrm{B}}\xspace}                 
                  
 \def\PD      {\ensuremath{\mathrm{D}}\xspace}

 \def\PK      {\ensuremath{\mathrm{K}}\xspace}

 \def\Pb      {\ensuremath{\mathrm{b}}\xspace}                 
 \def\Pc      {\ensuremath{\mathrm{c}}\xspace}

 \def\Pi      {\ensuremath{\mathrm{i}}\xspace}

 \def\Ps      {\ensuremath{\mathrm{s}}\xspace}

 \def\thebaroffset{0.0em}
}
{

 \def\Pmu         {\ensuremath{\mu}\xspace}

 \def\Ppi         {\ensuremath{\pi}\xspace}

 \mathchardef\PDelta="7101
 \mathchardef\PXi="7104
 \mathchardef\PLambda="7103
 \mathchardef\PSigma="7106
 \mathchardef\POmega="710A
 \mathchardef\PUpsilon="7107
                  
 \def\PB      {\ensuremath{B}\xspace}                 
                  
 \def\PD      {\ensuremath{D}\xspace}

 \def\PK      {\ensuremath{K}\xspace}

 \def\Pb      {\ensuremath{b}\xspace}                 
 \def\Pc      {\ensuremath{c}\xspace}

 \def\Pi      {\ensuremath{i}\xspace}

 \def\Ps      {\ensuremath{s}\xspace}

 \def\thebaroffset{0.18em}
}
\newcommand{\offsetoverline}[2][\thebaroffset]{\kern #1\overline{\kern -#1 #2}}%

\makeatletter
\ifcase \@ptsize \relax
  \newcommand{\miniscule}{\@setfontsize\miniscule{4}{5}}
\or
  \newcommand{\miniscule}{\@setfontsize\miniscule{5}{6}}
\or
  \newcommand{\miniscule}{\@setfontsize\miniscule{5}{6}}
\fi
\makeatother

\DeclareRobustCommand{\optbar}[1]{\shortstack{{\miniscule (\rule[.5ex]{1.25em}{.18mm})}
  \\ [-.7ex] $#1$}}

\def\mup        {{\ensuremath{\Pmu^+}}\xspace}

\def\mumu       {{\ensuremath{\Pmu^+\Pmu^-}}\xspace}

\def\squark    {{\ensuremath{\Ps}}\xspace}

\def\cquark    {{\ensuremath{\Pc}}\xspace}

\def\bquark    {{\ensuremath{\Pb}}\xspace}

\def\pion   {{\ensuremath{\Ppi}}\xspace}
\def\piz    {{\ensuremath{\pion^0}}\xspace}

\def\pim    {{\ensuremath{\pion^-}}\xspace}

\def\kaon    {{\ensuremath{\PK}}\xspace}
\def\Kbar    {{\ensuremath{\offsetoverline{\PK}}}\xspace}

\def\KorKbar {\kern \thebaroffset\optbar{\kern -\thebaroffset \PK}{}\xspace}
\def\Kz      {{\ensuremath{\kaon^0}}\xspace}
\def\Kzb     {{\ensuremath{\Kbar{}^0}}\xspace}

\def\KS      {{\ensuremath{\kaon^0_{\mathrm{S}}}}\xspace}

\def\KL      {{\ensuremath{\kaon^0_{\mathrm{L}}}}\xspace}

\def\DorDbar {\kern \thebaroffset\optbar{\kern -\thebaroffset \PD}\xspace}

\def\B       {{\ensuremath{\PB}}\xspace}

\def\BorBbar {\kern \thebaroffset\optbar{\kern -\thebaroffset \PB}\xspace}

\def\Bd      {{\ensuremath{\B^0}}\xspace}

\def\BdorBdbar {\kern \thebaroffset\optbar{\kern -\thebaroffset \Bd}\xspace}

\def\Bs      {{\ensuremath{\B^0_\squark}}\xspace}

\def\BsorBsbar {\kern \thebaroffset\optbar{\kern -\thebaroffset \Bs}\xspace}

\def\Y#1S{\ensuremath{\PUpsilon{(#1S)}}\xspace}

\def\Lz          {{\ensuremath{\PLambda}}\xspace}

\def\LorLbar     {\kern \thebaroffset\optbar{\kern -\thebaroffset \PLambda}\xspace}

\def\Sigmares    {{\ensuremath{\PSigma}}\xspace}

\def\BF         {{\ensuremath{\mathcal{B}}}\xspace}
\def\BR         {\BF}

\newcommand{\decay}[2]{\ensuremath{#1\!\to #2}\xspace} 

\def\to                 {\ensuremath{\rightarrow}\xspace}

\def\CP                {{\ensuremath{C\!P}}\xspace}

\def\AT#1     {\ensuremath{A_{\mathrm{T}}^{#1}}\xspace}           

\def\C#1      {\ensuremath{\mathcal{C}_{#1}}\xspace}                       
\def\Cp#1     {\ensuremath{\mathcal{C}_{#1}^{'}}\xspace}                    
\def\Ceff#1   {\ensuremath{\mathcal{C}_{#1}^{\mathrm{(eff)}}}\xspace}        
\def\Cpeff#1  {\ensuremath{\mathcal{C}_{#1}^{'\mathrm{(eff)}}}\xspace}       
\def\Ope#1    {\ensuremath{\mathcal{O}_{#1}}\xspace}                       
\def\Opep#1   {\ensuremath{\mathcal{O}_{#1}^{'}}\xspace}                    

\newcommand{\nospaceunit}[1]{\ensuremath{\text{#1}}}       
\newcommand{\aunit}[1]{\ensuremath{\text{\,#1}}}       

\newcommand{\tev}{\aunit{Te\kern -0.1em V}\xspace}
\newcommand{\gev}{\aunit{Ge\kern -0.1em V}\xspace}
\newcommand{\mev}{\aunit{Me\kern -0.1em V}\xspace}
\newcommand{\kev}{\aunit{ke\kern -0.1em V}\xspace}
\newcommand{\ev}{\aunit{e\kern -0.1em V}\xspace}
\newcommand{\mevc}{\ensuremath{\aunit{Me\kern -0.1em V\!/}c}\xspace}
\newcommand{\gevc}{\ensuremath{\aunit{Ge\kern -0.1em V\!/}c}\xspace}
\newcommand{\mevcc}{\ensuremath{\aunit{Me\kern -0.1em V\!/}c^2}\xspace}
\newcommand{\gevcc}{\ensuremath{\aunit{Ge\kern -0.1em V\!/}c^2}\xspace}

\def\mum  {\ensuremath{\,\upmu\nospaceunit{m}}\xspace}

\def\fb   {\ensuremath{\aunit{fb}}\xspace}
\def\invfb   {\ensuremath{\fb^{-1}}\xspace}

\newcommand{\chisq}{\ensuremath{\chi^2}\xspace}

\def\gsim{{~\raise.15em\hbox{$>$}\kern-.85em
          \lower.35em\hbox{$\sim$}~}\xspace}
\def\lsim{{~\raise.15em\hbox{$<$}\kern-.85em
          \lower.35em\hbox{$\sim$}~}\xspace}

\def\pt         {\ensuremath{p_{\mathrm{T}}}\xspace}

\def\ptot       {\ensuremath{p}\xspace}

\def\tell1  {TELL1\xspace}
\def\ukl1   {UKL1\xspace}

\usepackage{cite} 
\usepackage{mciteplus}

\newcommand{\mathcmd}[1]{\ensuremath{#1}\xspace}

\newcommand{\Ks}{\KS}
\newcommand{\Kl}{\KL}

\newcommand{\Ksmumu}{\ensuremath{\Ks\to\mu^+\mu^-}\xspace}

\newcommand{\Klmumu}{\ensuremath{\Kl\to\mu^+\mu^-}\xspace}
\newcommand{\Kspipi}{\ensuremath{\Ks\to\pi^+\pi^-}\xspace}

\newcommand{\Jpsimumu}{\ensuremath{J/\psi\to \mu^+\mu^-}\xspace}

\newcommand{\BRof}[1]{\ensuremath{{\cal B}(#1)}\xspace}

\newcommand{\IP}{\ensuremath{{\rm IP}}\xspace}

\newcommand{\IPchisq}{\mathcmd{\chisq_\IP}}

\newcommand{\swave}{{\it S-wave}\xspace}
\newcommand{\pwave}{{\it P-wave}\xspace}

\newcommand{\figref}[1]{Fig.~\ref{#1}}

\newcommand{\refapp}[1]{Appendix~\ref{#1}}

\usepackage{subfig}

\begin{document}

\renewcommand{\thefootnote}{\fnsymbol{footnote}}
\setcounter{footnote}{1}


\begin{titlepage}
  \pagenumbering{roman}

  \vspace*{-1.5cm}
  \centerline{\large EUROPEAN ORGANIZATION FOR NUCLEAR RESEARCH (CERN)}
  \vspace*{1.5cm}
  \noindent
  \begin{tabular*}{\linewidth}{lc@{\extracolsep{\fill}}r@{\extracolsep{0pt}}}
    \vspace*{-1.5cm}\mbox{\!\!\!\includegraphics[width=.14\textwidth]{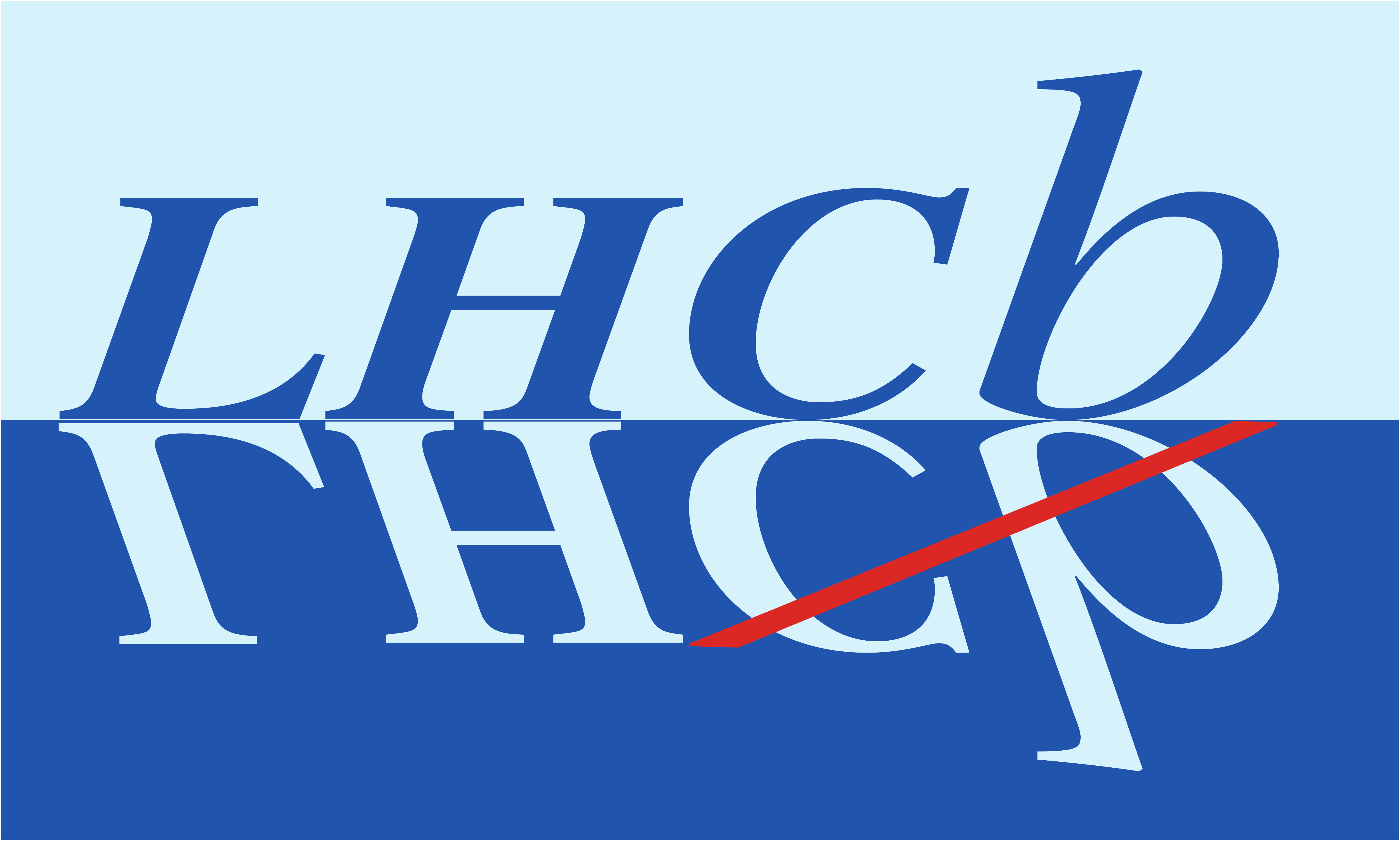}} & &%
    \\
    & & CERN-EP-2019-285 \\  
    & & LHCb-PAPER-2019-038 \\  
    & & December 2, 2020 \\ 
    & & \\
  \end{tabular*}

  \vspace*{4.0cm}

  {\normalfont\bfseries\boldmath\huge
    \begin{center}
      \papertitle
    \end{center}
  }

  \vspace*{2.0cm}

  \begin{center}
    \paperauthors\footnote{Authors are listed at the end of this paper.}
  \end{center}

  \vspace{\fill}

  \begin{abstract}
  \noindent 
    A search for the decay $K^0_{\mathrm{S}}\rightarrow\mu^{+}\mu^{-}$ is performed using  proton-proton collision data, corresponding to an integrated luminosity of
    $5.6\invfb$ and collected with the LHCb experiment during 2016, 2017 and 2018 at a center-of-mass energy of $13\,\mathrm{TeV}$.
    The observed signal yield is consistent with zero, yielding an upper limit of \mbox{${\cal B}(K^0_{\mathrm{S}}\rightarrow\mu^{+}\mu^{-}) < 2.2 \times 10^{-10}$} at 90\% CL. The limit reduces to \mbox{${\cal B}(K^0_{\mathrm{S}}\rightarrow\mu^{+}\mu^{-}) < 2.1 \times 10^{-10}$} at 90\% CL once combined with the result from data taken in 2011 and 2012.
  \end{abstract}

  \vspace*{2.0cm}

  \begin{center}
    Published in Phys.~Rev.~Lett.~\textbf{125},~231801~(2020)
  \end{center}

  \vspace{\fill}

  {\footnotesize
    \centerline{\copyright~\papercopyright. \href{\paperlicenceurl}{\paperlicence}.}}
  \vspace*{2mm}

\end{titlepage}


\newpage
\setcounter{page}{2}
\mbox{~}
%
%
%
%

\cleardoublepage


\renewcommand{\thefootnote}{\arabic{footnote}}
\setcounter{footnote}{0}



\pagestyle{plain} 
\setcounter{page}{1}
\pagenumbering{arabic}

\begin{figure}[b]
  \centering
  \subfloat{\includegraphics[scale=0.7]{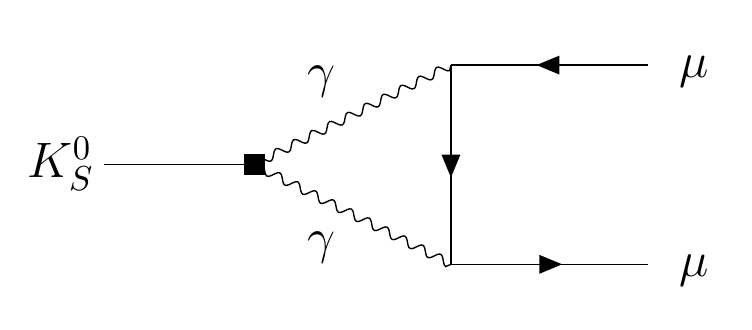}}\\
  \subfloat{\includegraphics[scale=0.7]{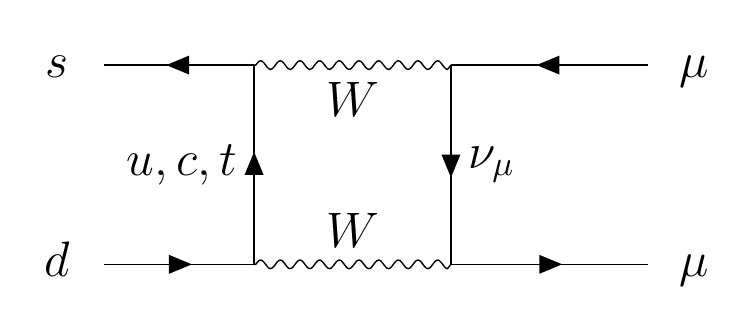}}
  \subfloat{\includegraphics[scale=0.7]{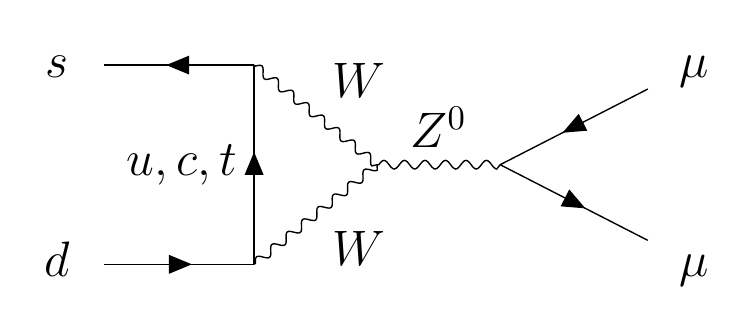}}
  \subfloat{\includegraphics[scale=0.7]{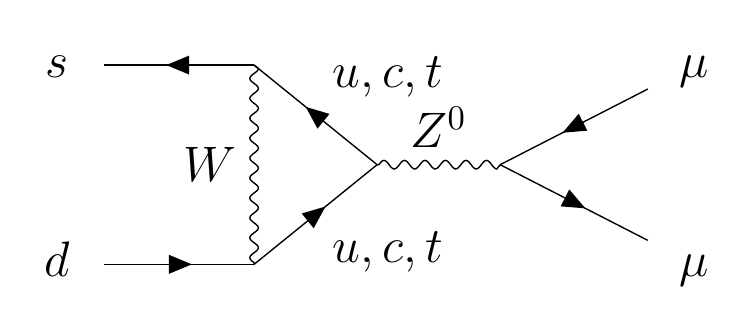}}
  \caption{%
    Diagrams representing SM contributions to the \Ksmumu decay amplitude: (top) long-distance contribution, generated by two intermediate photons, and
    (bottom) short-distance contributions.
  }
  \label{fig:feynman-diagrams}
\end{figure}

The decay \Ksmumu is a flavor-changing neutral current (FCNC) process which has not been observed yet.
In the standard model (SM), this decay is highly suppressed~\cite{Ecker:1991ru,Isidori:2003ts}, with an expected branching fraction \mbox{$\BRof\Ksmumu_{\rm SM} = (5.18 \pm 1.50_{\rm LD} \pm 0.02_{\rm SD}) \times 10^{-12}$}~\cite{DAmbrosio:2017klp}.
The uncertainties with subscripts LD and SD relate to long-distance and short-distance effects, respectively.
The main contributions to the \Ksmumu decay amplitude are illustrated in Fig.~\ref{fig:feynman-diagrams}.
The related channel \Klmumu is predicted in the SM to occur with a branching fraction
$\BRof\Klmumu_{\rm SM} = (6.85 \pm 0.80_{\rm LD} \pm 0.06_{\rm SD} ) \times 10^{-9}$ or
$\BRof\Klmumu_{\rm SM} =(8.11 \pm 1.49_{\rm LD} \pm 0.13_{\rm SD} ) \times 10^{-9}$
for an (unknown) positive or a negative relative sign of the \decay{\KL}{\gamma\gamma} amplitude~\cite{D'Ambrosio:1994ae}, respectively.
These predictions are in good agreement with the experimental world average \mbox{$\BR(\decay{\KL}{\mumu}) = (6.84 \pm 0.11) \times 10^{-9}$}~\cite{PDG2018}, based on Refs.~\cite{Ambrose:2000gj,Akagi:1994bb,Heinson:1994kq}.
Both the \Ks and the \KL decay amplitudes are dominated by LD contributions in the SM.
The large difference between the two branching fractions is due to the \swave component, which is charge-parity (\CP) violating and \CP conserving for the \Ks and \KL modes, respectively.
In the \Ks case, the \CP-conserving long-distance contribution can only proceed through the \pwave, and the \CP-violating short distance component in the SM is even more suppressed.

Because of the strong suppression of the SM decay amplitude, dynamics beyond the standard model (BSM) can induce large deviations of \BRof\Ksmumu with respect to the SM prediction.
This has been shown to be the case in SUSY scenarios~\cite{Chobanova:2017rkj} as well as in leptoquark models~\cite{Dorsner:2011ai,Bobeth:2017ecx}.
The current best limit, ${\cal B}(\Ksmumu) < 0.8 \times 10^{-9}$ at 90\% confidence level (C.L.), was set by LHCb~\cite{LHCb-PAPER-2017-009} with the data collected during Run~1 (2011--2012).

In this letter, a significantly improved limit is presented.
Results are based on proton-proton ($pp$) collision data collected with the LHCb detector at a center-of-mass energy of $13\tev$ during 2016, 2017, and 2018 (Run~2), corresponding to an integrated luminosity of $5.6\invfb$.
This measurement benefits from the huge \Ks production cross section at the LHC of approximately $0.6$ b at a center-of-mass energy of $13\tev$~\cite{Junior:2018odx}, and from the forward geometry of the vertex detector of LHCb since \Ks mesons are predominantly produced at low angles with respect to the beam pipe.
A major improvement with respect to the previous analysis is achieved by employing dedicated software triggers that were not present in Run~1.
These new triggers were included from the start of 2016 data taking, so data from 2015 is not used, due to a lower trigger efficiency and integrated luminosity.
While the analysis strategy closely follows what was done for Run~1, the event reconstruction and selection have been improved.

The LHCb detector~\cite{LHCb-DP-2012-002,LHCb-DP-2014-002} is a single-arm forward spectrometer covering the pseudorapidity range $2 < \eta < 5$, designed for
the study of particles containing \bquark or \cquark quarks.
The detector includes a high-precision tracking system consisting of a silicon-strip vertex
detector (VELO) surrounding the $pp$ interaction region, a large-area silicon-strip detector
located upstream of a dipole magnet with a bending power of about
$4{\mathrm{\,Tm}}$, and three stations of silicon-strip detectors and straw
drift tubes placed downstream of the magnet.
The tracking system provides a measurement of the momentum, \ptot, of charged
particles with a relative uncertainty that varies from 0.5\% at low momentum to
1.0\% at 200\gevc.
The minimum distance of a track to a proton-proton collision vertex (PV),
the impact parameter (IP), is measured with a resolution of $(15+29/\pt)\mum$,
where \pt is the component of the momentum transverse to the beam axis, in \gevc.
Different types of charged hadrons are distinguished using information from two
ring-imaging Cherenkov (RICH) detectors.
Photons,
electrons and hadrons are identified by a calorimeter system consisting of
scintillating-pad and preshower detectors, an electromagnetic and a
hadronic calorimeter.
Muons are identified by a system composed of alternating
layers of iron and multiwire proportional
chambers.
In addition, information from the tracking system, the calorimeter system and the RICH detectors is used to further improve the muon identification.

Events are first required to pass a hardware-trigger selection~\cite{LHCb-DP-2012-004}, based on information from the calorimeter and the muon system, relying on high-\pt signatures.
Subsequently, a full event reconstruction is applied in a two-step software selection.
In the previous analysis, the search was limited by a muon \pt threshold of approximately $1.8\gevc$.
In Run~2, a new tracking method was included, in order to improve the reconstruction of muons with low transverse momentum.
By using the information from the muon chambers at early stages in the reconstruction chain, a drastic reduction of the number of tracks to be processed by the most time-consuming reconstruction algorithms is achieved.
This new reconstruction method allowed to reduce the \pt muon threshold to $80\mevc$.
In addition, a dedicated software trigger selection was developed, using the aforementioned reconstruction method, fully covering the dimuon invariant mass spectra of many strange decays, including \Ksmumu.
This translates into an increase of the trigger efficiency for \decay{\Ks}{\mumu} of about an order of magnitude with respect to Run~1~\cite{Dettori:2297352}.
After the upgrade of the LHCb detector~\cite{LHCb-TDR-018}, the hardware trigger will no longer be present, allowing for further efficiency improvements.

The purity of the signal candidates and the evaluation of the systematic uncertainties depend on the hardware trigger requirements, so the full data sample is divided into two categories.
In the first category, referred to as triggered-independent-of-signal (TIS), events are triggered at the hardware stage independently of the trigger decision on the decay products of the signal candidate.
The second category, referred to as exclusively triggered-on-signal (XTOS), consists of events triggered at the hardware stage by the signal candidate decay products that are not contained in the TIS category~\cite{LHCb-2008-073}.
Both categories are required to fulfill the same software trigger requirements.

The measurement of the \Ksmumu branching fraction requires the normalization to the \Ks meson production rate, which is done using \Kspipi decays, given its abundance, its similar topology to \Ksmumu and its well-known branching fraction~\cite{PDG2018}.
Common off-line preselection criteria are applied to \Ksmumu and \Kspipi candidates in order to reduce many systematic effects in the efficiency ratio.
Candidate \Ksmumu (\Kspipi) decays are obtained from two tracks with opposite charge identified as muons (pions), forming a secondary vertex (SV) and with an invariant mass in the range \mbox{400--600\mevcc}.
Kaon candidates are required to decay inside the VELO, where the best \Ks invariant mass resolution is achieved. Approximately $22\%$ of \Ks mesons produced at the $pp$ interaction point decay within the acceptance of the VELO.
The \Ks candidate origin must be compatible with a PV, while its decay products should be inconsistent with originating from any PV.
The SV must be well detached from the PV by requiring the \Ks candidate decay time to be larger than 6\% of the known \Ks lifetime~\cite{PDG2018}.
Decays of \Lz baryons to $p \pim$, and the charge-conjugate counterpart, are suppressed by removing candidates close to the
expected elliptical kinematic regions in the Armenteros--Podolanski plane~\cite{armenteros} (The inclusion of charge-conjugate processes is implied throughout this paper, unless otherwise noted.). The corresponding loss in signal efficiency is negligible.
Muon tracks are required to have associated hits in the muon system~\cite{LHCb-DP-2013-001},
while pions from \Kspipi decays are required to be within the muon system acceptance.
The main background sources are random combinations of tracks, inelastic interactions with the detector material, and \Kspipi decays, where the two pions are misidentified as muons.
In doubly misidentified \Kspipi decays, the invariant mass of the kaon candidate is underestimated on average by $40\mevcc$, corresponding to ten times the dimuon invariant mass resolution in this energy regime.

Background from material interactions and random combinations of tracks is suppressed using two adaptive boosted decision tree (BDT)~\cite{AdaBoost,Breiman} algorithms based on the XGBoost library~\cite{XGBoost} and optimized for each trigger category.
Simulated \Ksmumu decays are used as a proxy for signal, and \Ksmumu candidates from data in the dimuon invariant mass region above $520\mevcc$ as a proxy for background.
Data from the left sideband are not considered since it is dominated by doubly misidentified \Kspipi decays.
Before the BDT training, the simulated \Ksmumu candidates are weighted using a gradient boost algorithm~\cite{Rogozhnikov:2016bdp} trained with \Kspipi candidates in simulation and data, to take into account small differences between data and simulation.
Since the background candidates used in the training are part of the fitted sample, the {\it k-folding} approach~\cite{scikit-learn} is applied to maximize the sample available without biasing the background estimate.
The BDT input variables are
the kaon candidate decay time and IP significance (\IPchisq), defined as the increase of the \chisq of the PV when considering the kaon candidate in the vertex fit;
the \IPchisq and the track-fit $\chi^2$ of each of the two tracks;
the distance of closest approach between the two tracks;
the cosine of the helicity angle;
the $\chi^2$ of the SV fit;
two SV isolation variables, defined as the difference in the $\chi^2$ in the vertex fit with
only the two final-state tracks and that obtained when adding the one or two nearest tracks; and
a VELO material veto variable~\cite{LHCb-DP-2018-002}.
The VELO material veto variable efficiently suppresses background originating from inelastic interactions with the VELO stations and RF foil which separates the VELO modules from the beam vacuum.
See~\refapp{app:supplemental} for more information about the material veto and the interactions with the VELO material.
A selection requirement is placed on the BDT, rejecting 99\% of the background with a signal efficiency of approximately 63\% for both trigger categories.

Another significant background source is \Klmumu decays, for which the LHCb detector has the efficiency suppressed by a factor of approximately $2.3\times 10^{-3}$ relative to \Ksmumu decays due to its longer lifetime.
Interference between \KS and \KL mesons is neglected since \Kz and \Kzb mesons are expected to be produced in equal amounts~\cite{DAmbrosio:2017klp} at the LHC.
Contributions from other background sources, such as $\Kz\to\mu^+\mu^-\gamma(\gamma)$, $\Sigmares^+\to p\mu^+\mu^-$, $K^{0,+}\to\pi^{0,+}\mu^+\mu^-$, $\Lz\to p\pi^-$, $\omega\to\piz\mu^{+}\mu^{-}$, $\eta\to\mu^{+}\mu^{-}\gamma$, as well as from $\KL\to \pi^{\pm}\mu^{\mp}\nu_{\mu}$ and $\Ks\to \pi^{\pm}\mu^{\mp}\nu_{\mu}$ decays, the latter recently discovered by the \kloeii Collaboration~\cite{Babusci:2019gqu}, are found to be negligible.

Candidates satisfying the preselection criteria are divided into twenty subsets: ten bins of the BDT response for each of the two trigger categories.
The BDT bins are chosen to have the same fraction of simulated signal candidates in each bin.
A dedicated muon identification boosted decision tree ($\mu$BDT) is used to suppress \Kspipi decays, whose performance can be consulted in Ref.~\cite{LHCb-PAPER-2017-009}.
The selection criterion on the $\mu$BDT is optimized and applied independently for each of the twenty categories.
The response of the muon identification is calibrated using \Jpsimumu decays, complemented with the use of $\Kz\to\pim\mup\nu_\mu$ decays due to the lower transverse momentum of the decay products.

\begin{figure}[tb]
  \begin{center}
    \includegraphics[width=0.6\textwidth]{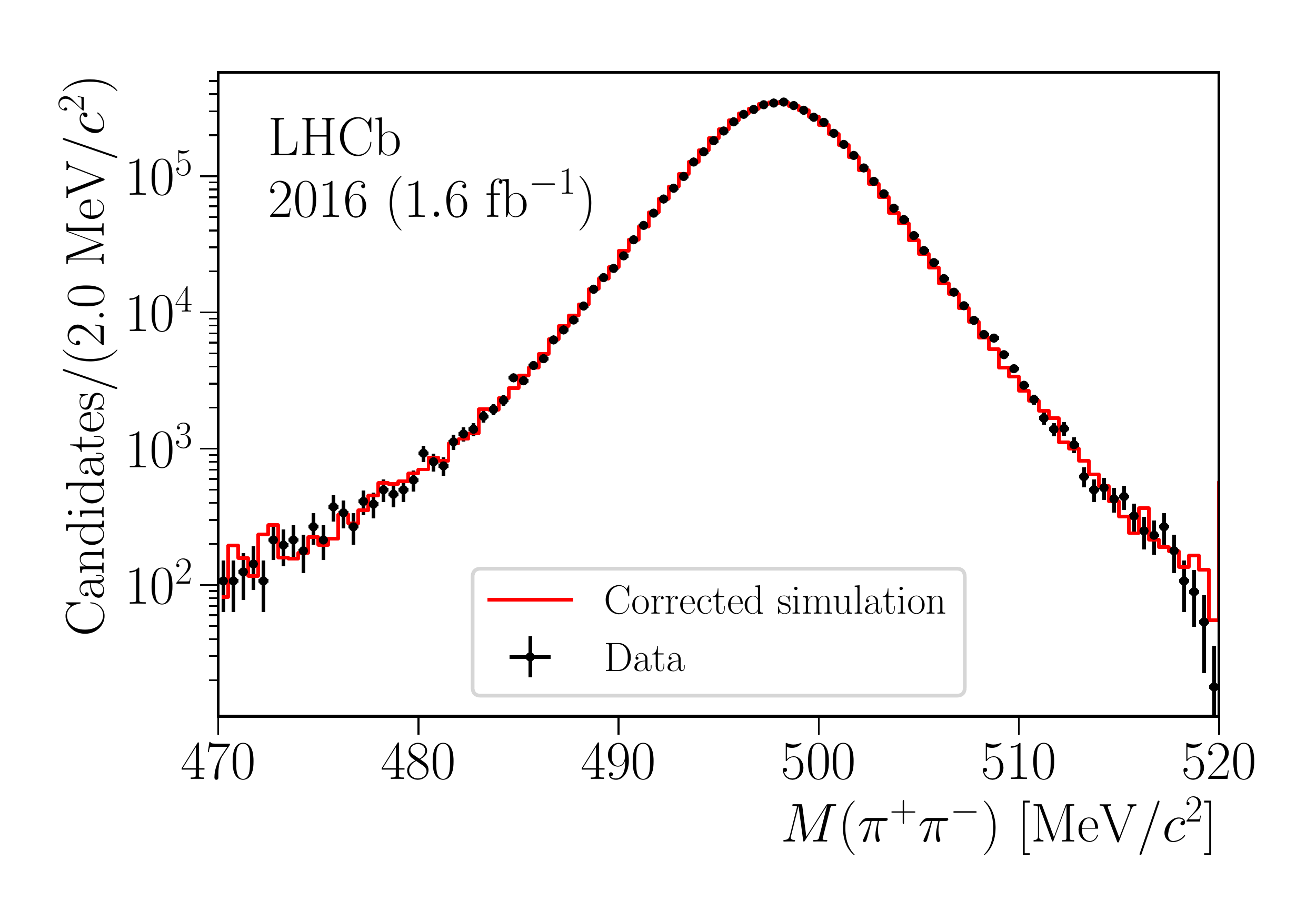}
    \caption{Invariant-mass distribution of \Kspipi candidates in 2016 trigger-unbiased data (points with error bars) and corrected simulation (solid histogram).
    The histogram of simulated candidates is normalized to  data.\label{fig:mass-calib}}
  \end{center}
\end{figure}

The \Ksmumu branching fraction is determined in an unbinned maximum-likelihood fit to the kaon candidate invariant mass in the range \mbox{480--595\mevcc}.
Taking into account the ratio of detection efficiencies, the signal yield is normalized to \Kspipi decays to cancel uncertainties due to the \KS cross section, luminosity, reconstruction, and partially due to selection criteria including the BDT binning.
The fit is performed simultaneously in the twenty data categories.
The contributions considered are \Ksmumu signal, modelled with a Hypatia distribution~\cite{Hypatia}; background from material interactions and random combination of tracks, described by an exponential function; the \Kspipi background, modelled with a power law distribution; and \Klmumu, described with the same probability density function as the \Ksmumu decay.
All yields are free to vary in the fit.
Because of the low level of background from material interactions and random combination of tracks, the slope of the exponential function is left to change sign when constructing the profile likelihood.
A Gaussian constraint is applied to the yield of the \Klmumu component, based on its known branching fraction~\cite{PDG2018} and on the efficiency ratios between \Klmumu and \Ksmumu.
Additional Gaussian constraints are applied to the efficiency ratios between \Ksmumu and \Kspipi, accounting for the systematic uncertainties.
An independent sample of \Kspipi decays obtained from a trigger-unbiased sample is used to calibrate the \KS invariant mass peak position and resolution parameters (see \figref{fig:mass-calib}).
It is also used to correct the simulation to obtain the efficiencies of the signal and the normalization channel.

The yield of $\Kz\to\pim\mup\nu_\mu$ decays as a function of the data taking period is also used to evaluate the variation of the total efficiency with time,
mostly caused by changes in the thresholds of the hardware trigger.
The obtained single-event sensitivity is $(3.0\pm0.6) \times 10^{-12}$, meaning that approximately two \Ksmumu and five \Klmumu signal decays are expected to be present in the data set, using the SM prediction for the branching fractions, and also taking into account the \Klmumu detection suppression of $2.3 \times 10^{-3}$.

\begin{figure}[tb]
  \begin{center}
    {\includegraphics[width=0.49\textwidth]{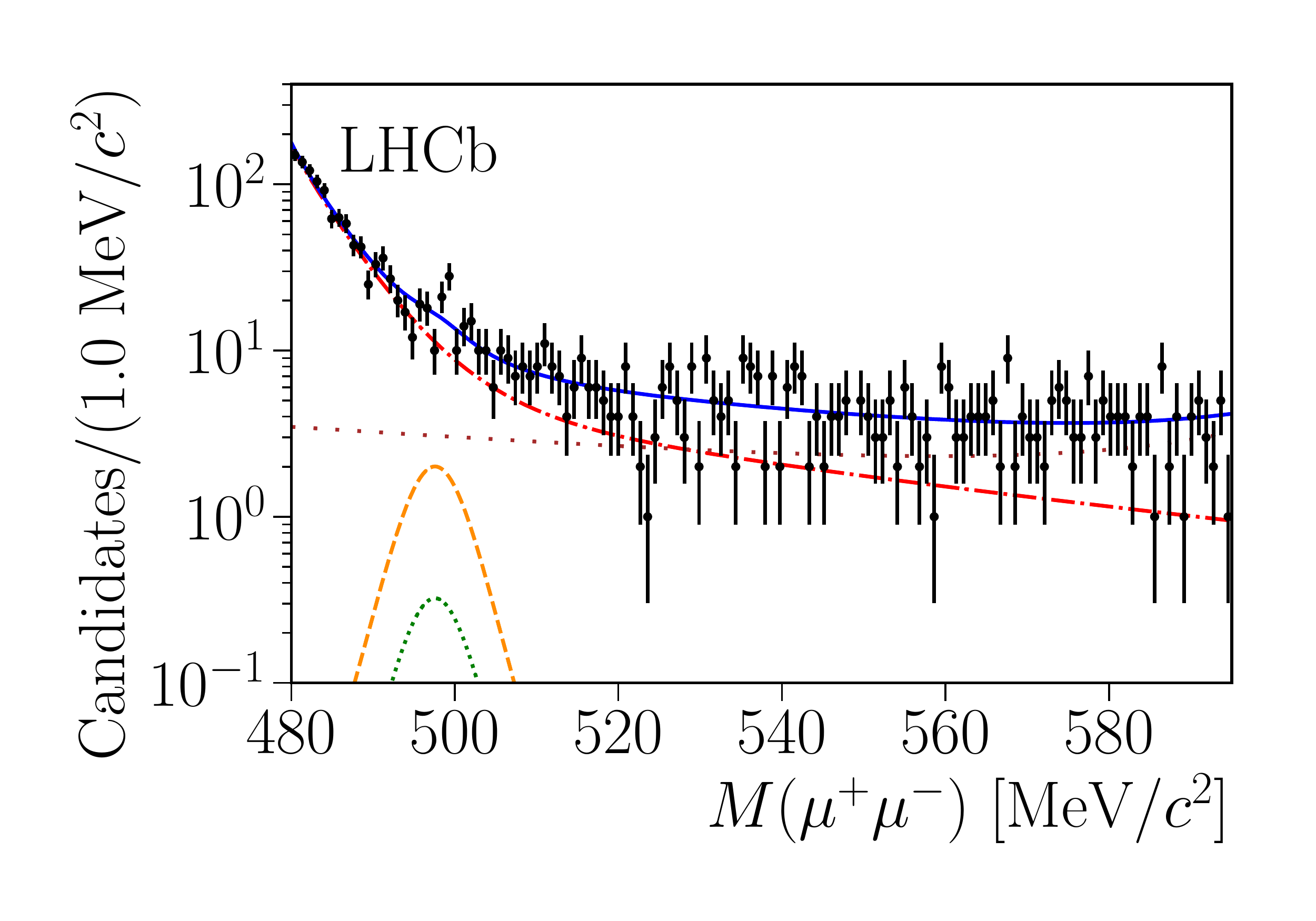}}
    {\includegraphics[width=0.49\textwidth]{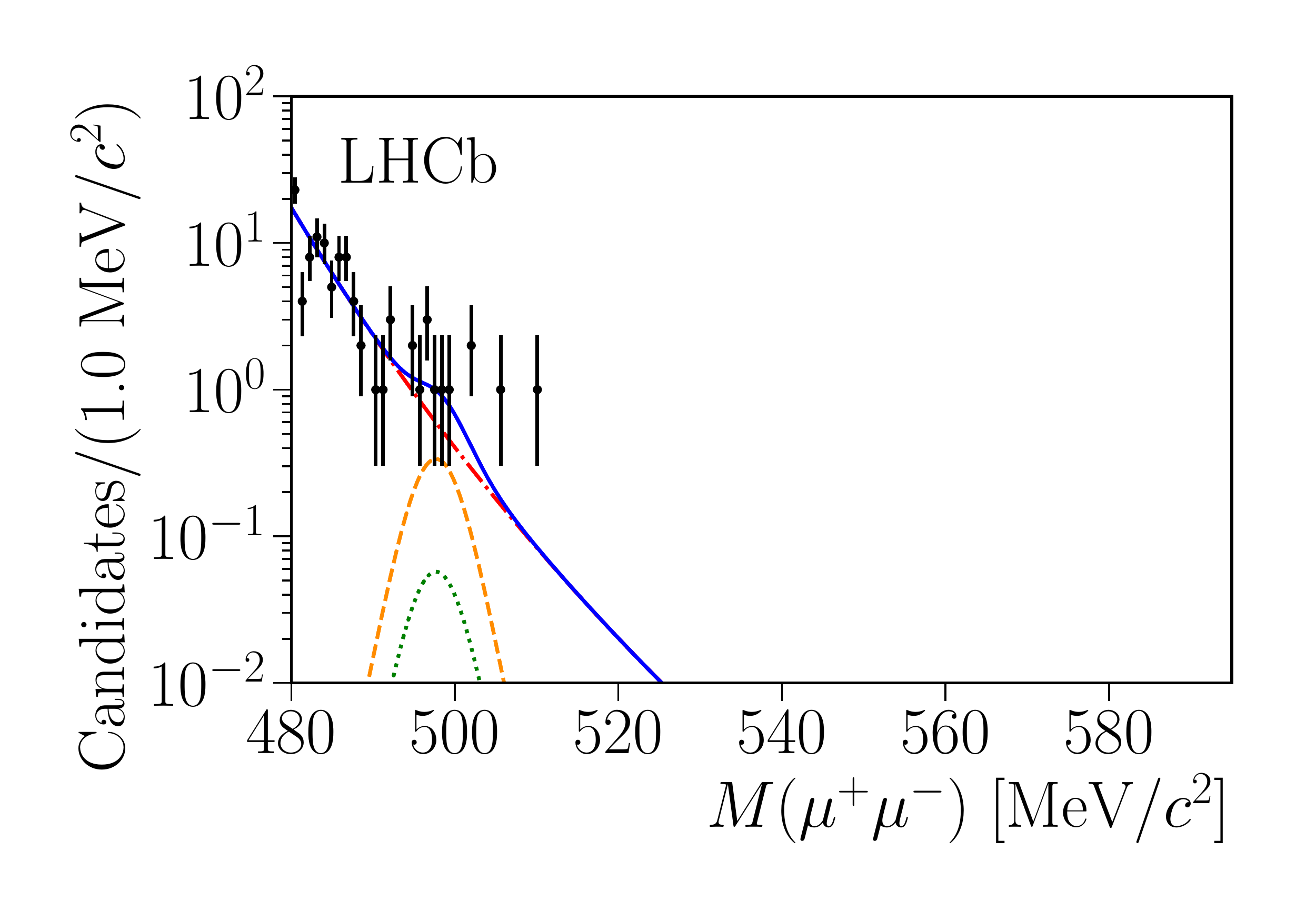}}

    {\includegraphics[width=0.49\textwidth]{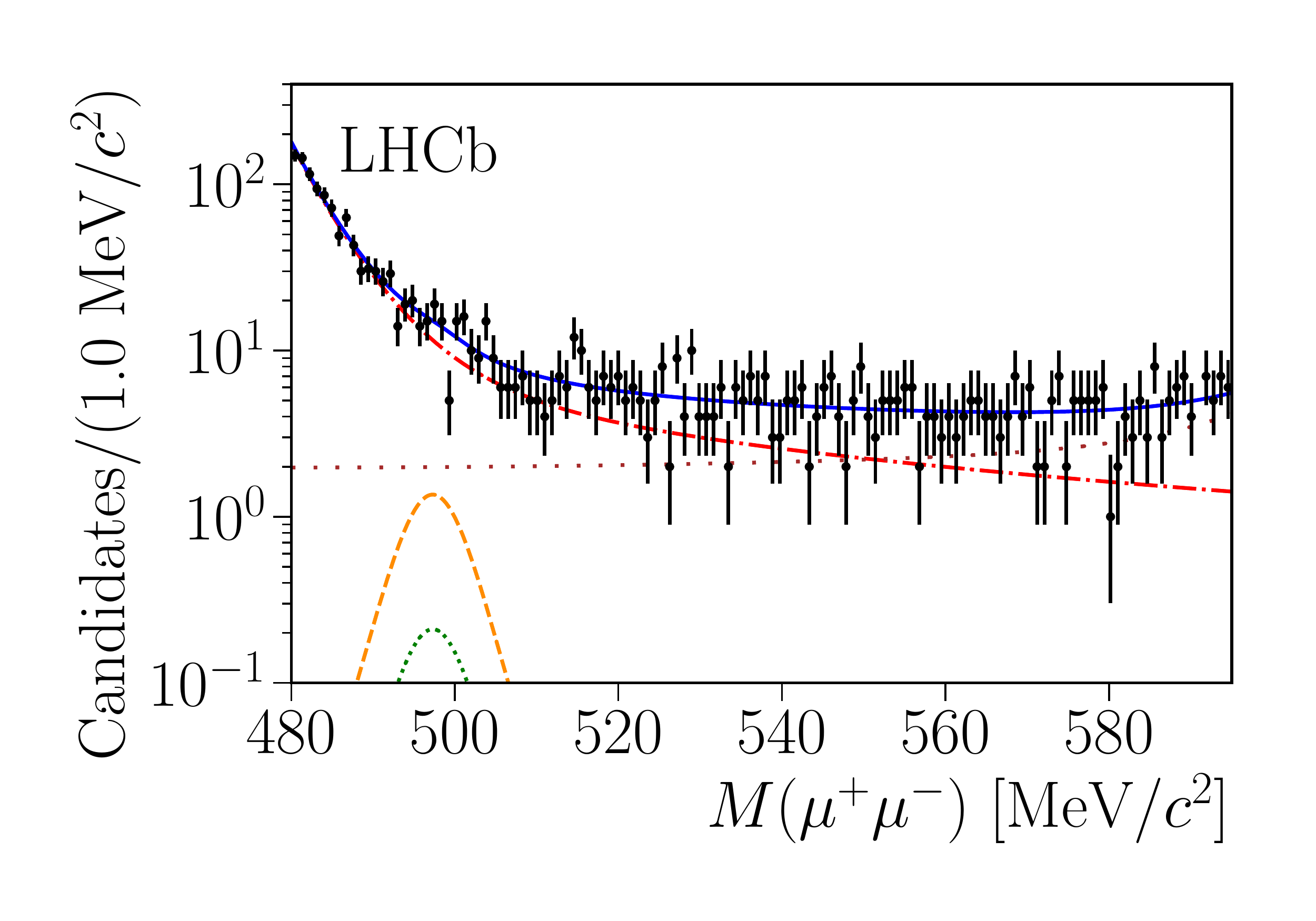}}
    {\includegraphics[width=0.49\textwidth]{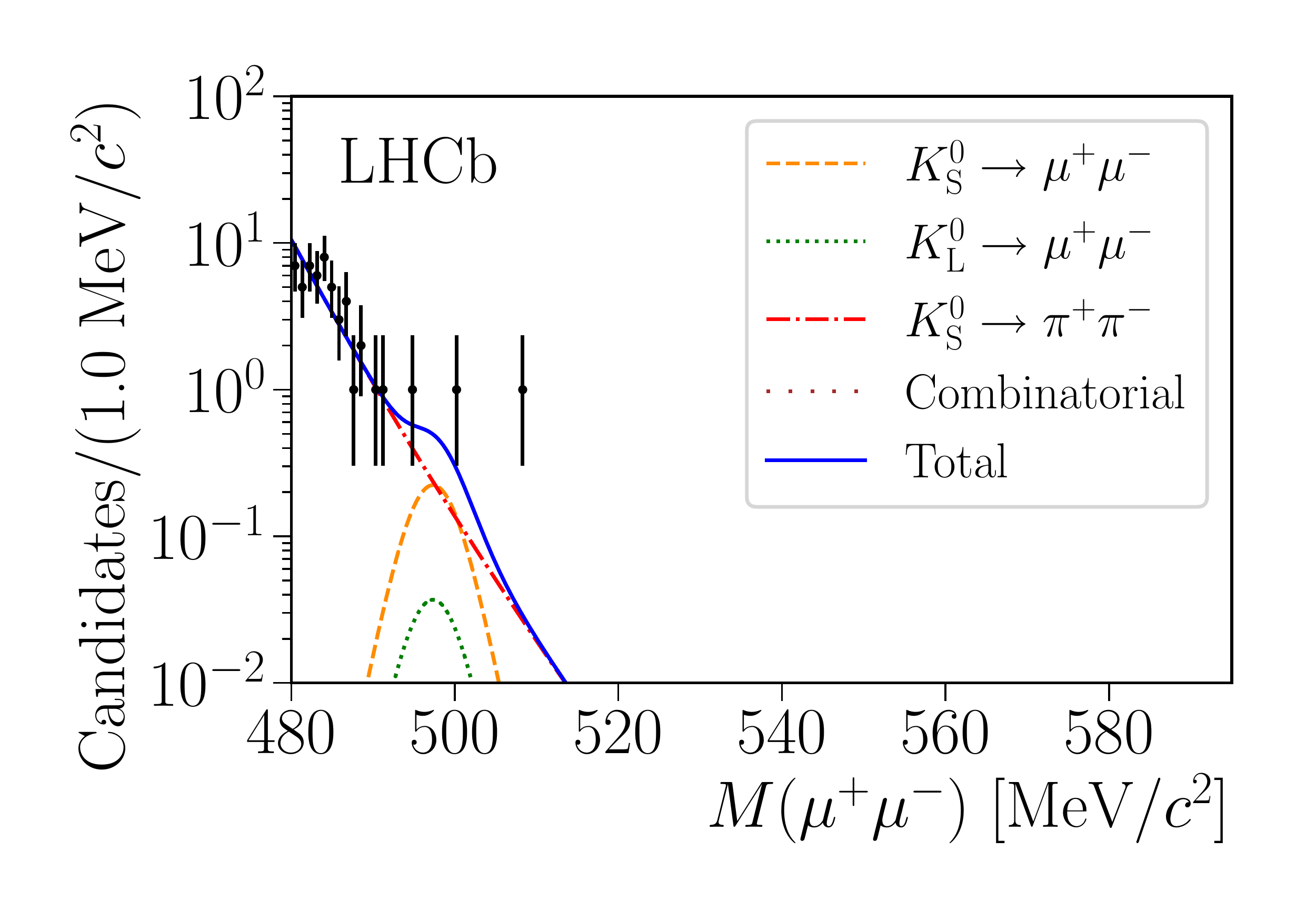}}

    \caption{%
      Projection of the fit to the dimuon invariant mass distribution for the (top left) TIS and (bottom left) XTOS trigger categories.
      The plots on the right correspond to the projection of the fit in the BDT bins with the highest signal-to-background ratio for the (top right) TIS and (bottom right) XTOS trigger categories.
      The dashed orange line shows the signal contribution, the dotted green line the \mbox{\Klmumu} contribution, the dash dotted red line the \Kspipi contribution, the loosely dotted brown line the background from random combination of tracks and material interactions, and the solid blue line the total probability density function.
      For clarity, empty bins are not shown.
    }
    \label{fig:fit}
  \end{center}
\end{figure}

Various sources of systematic uncertainty are taken into account.
The main sources are the determination of the trigger efficiency, yielding a systematic uncertainty of 11\% for the hardware trigger and 13\% for the software trigger; data-simulation differences in the muon identification, with systematic uncertainties varying between $4\%$ and $12\%$, depending on the trigger category and BDT bin; and the correction applied on simulation, evaluated to be $6\%$.
Other sources, like the efficiency ratio between the signal and normalization modes, the BDT response due to changes in the experimental conditions, and the uncertainty on the \Kspipi branching fraction are found to be smaller than $5\%$.
The total systematic uncertainty is between 19\% and 23\%, depending on the trigger category and the BDT bin.
It tends to be lower in the TIS trigger category and higher in lower BDT bins, which have lower signal-to-background ratio, due to the stronger muon identification requirements for the lower bins and the bigger systematic uncertainty for the XTOS trigger efficiency.
The  systematic uncertainties are taken into account as Gaussian constraints in the fit to the data.

\begin{figure}[t]
  \begin{center}
    \includegraphics[width=0.6\textwidth]{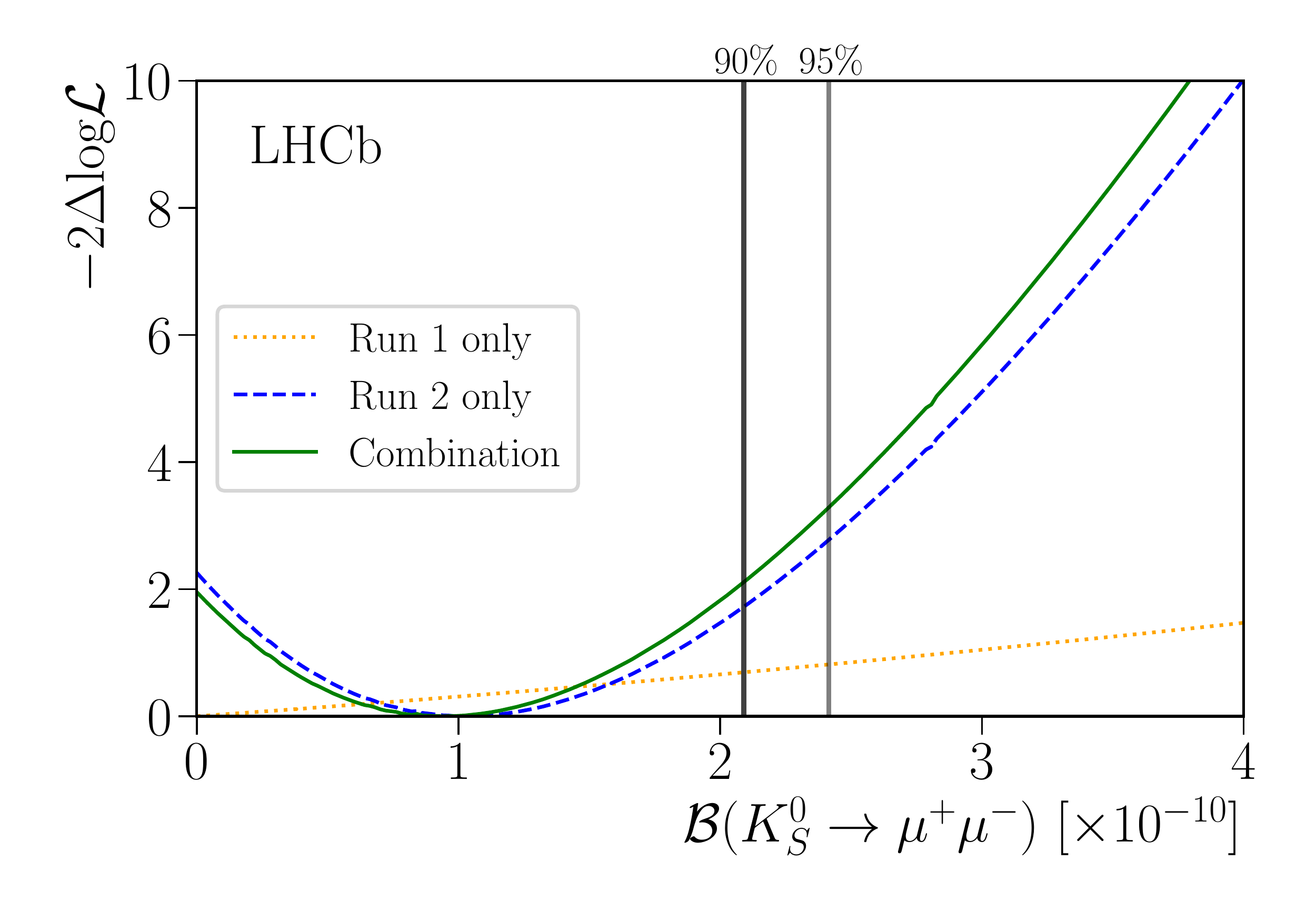}
    \caption{ Evaluation of $-2\Delta \log {\cal L}$, where ${\cal L}$ is the likelihood of the fit model, as a function of \BRof\Ksmumu.
    The dotted orange line corresponds to the Run~1 result, the dashed blue line to the Run~2 result, and the solid green line shows the combination.
    The two vertical lines show the location of the upper limit of the combined result at 90\% and 95\% confidence level. \label{fig:profLL}}
  \end{center}
\end{figure}

The expected significance for a SM signal is \(0.1\sigma\), and the expected upper limit is evaluated to be \(1.2(1.5)\times10^{-10}\) at 90(95)\% C.L.
The fit shows no evidence for \Ksmumu decays (see~\figref{fig:fit}), with a total yield of $34\pm 23$ signal candidates.
The signal yield is consistent with zero for all the BDT bins of the two trigger categories.
The significance with respect to the background-only hypothesis is $1.5\sigma$ ($1.4 \sigma$ when combined with Run~1 data).
An upper limit on the branching fraction is obtained by integrating the profile likelihood multiplied by a flat prior in the positive branching fraction domain, yielding $2.2\,(2.6)\times 10^{-10}$ at 90(95)\% C.L.
The likelihood is combined with the Run~1 result, obtaining a limit of $2.1\,(2.4)\times 10^{-10}$ at 90(95)\% C.L.
A log-likelihood interval of one standard deviation ($-2\Delta \log{\cal L} = 1$) from the Run~2 data set yields
\mbox{$\BRof\Ksmumu = 1.0_{-0.7}^{+0.8}\times 10^{-10}$}.
Combined with Run~1 it yields \mbox{$\BRof\Ksmumu = 0.9_{-0.6}^{+0.7}\times 10^{-10}$}.
The profile likelihoods are shown in \figref{fig:profLL}.

In summary, a search for the rare decay \Ksmumu has  been  performed  on  a LHCb data set of about $8.6\invfb$.
The obtained results  supersede  those  of  our  previous  publications~\cite{LHCb-PAPER-2017-009,LHCb-PAPER-2012-023}.
The data are consistent both with the background-only hypothesis and the combined background and SM signal expectation at the  $1.4 \sigma$ and $1.3\sigma$ level, respectively.
The most stringent upper limit on the \Ksmumu branching fraction to date of $2.1\,(2.4)\times 10^{-10}$ at 90(95)\% confidence level is set, improving the previous best limit by a factor of four.
\vskip 1em

\section*{Acknowledgements}
%
%
\noindent We would like to thank M. Moulson, J. Martin Camalich, and G. D'Ambrosio for fruitful discussions. 
We express our gratitude to our colleagues in the CERN
accelerator departments for the excellent performance of the LHC. We
thank the technical and administrative staff at the LHCb
institutes.
We acknowledge support from CERN and from the national agencies:
CAPES, CNPq, FAPERJ and FINEP (Brazil); 
MOST and NSFC (China); 
CNRS/IN2P3 (France); 
BMBF, DFG and MPG (Germany); 
INFN (Italy); 
NWO (Netherlands); 
MNiSW and NCN (Poland); 
MEN/IFA (Romania); 
MSHE (Russia); 
MinECo (Spain); 
SNSF and SER (Switzerland); 
NASU (Ukraine); 
STFC (United Kingdom); 
DOE NP and NSF (USA).
We acknowledge the computing resources that are provided by CERN, IN2P3
(France), KIT and DESY (Germany), INFN (Italy), SURF (Netherlands),
PIC (Spain), GridPP (United Kingdom), RRCKI and Yandex
LLC (Russia), CSCS (Switzerland), IFIN-HH (Romania), CBPF (Brazil),
PL-GRID (Poland) and OSC (USA).
We are indebted to the communities behind the multiple open-source
software packages on which we depend.
Individual groups or members have received support from
AvH Foundation (Germany);
EPLANET, Marie Sk\l{}odowska-Curie Actions and ERC (European Union);
ANR, Labex P2IO and OCEVU, and R\'{e}gion Auvergne-Rh\^{o}ne-Alpes (France);
Key Research Program of Frontier Sciences of CAS, CAS PIFI, and the Thousand Talents Program (China);
RFBR, RSF and Yandex LLC (Russia);
GVA, XuntaGal and GENCAT (Spain);
the Royal Society
and the Leverhulme Trust (United Kingdom).

\clearpage
\appendix
\section{Supplemental material}
\label{app:supplemental}

\begin{figure}[h]
  \centering
  \subfloat{\includegraphics[width=0.5\textwidth]{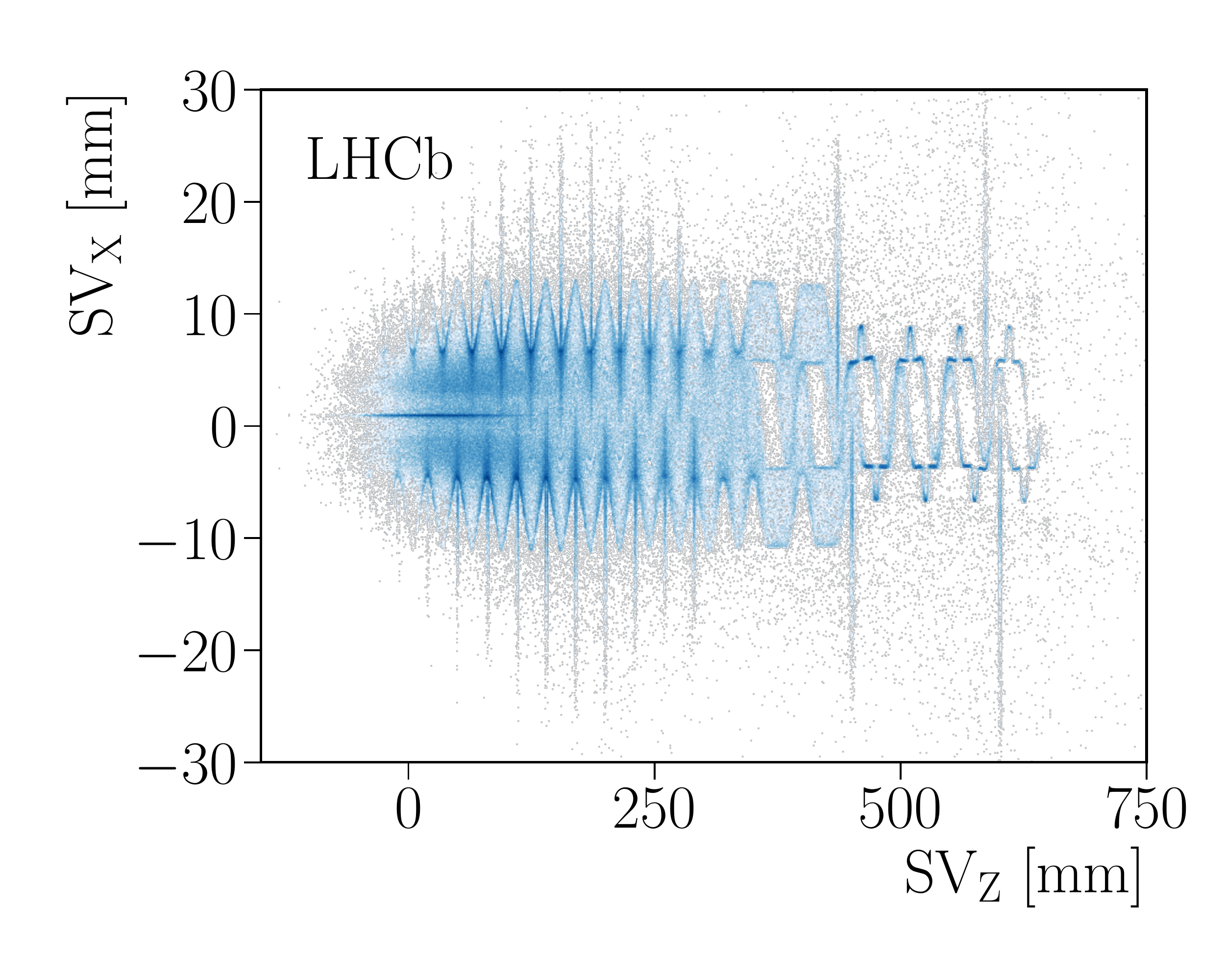}}
  \subfloat{\includegraphics[width=0.5\textwidth]{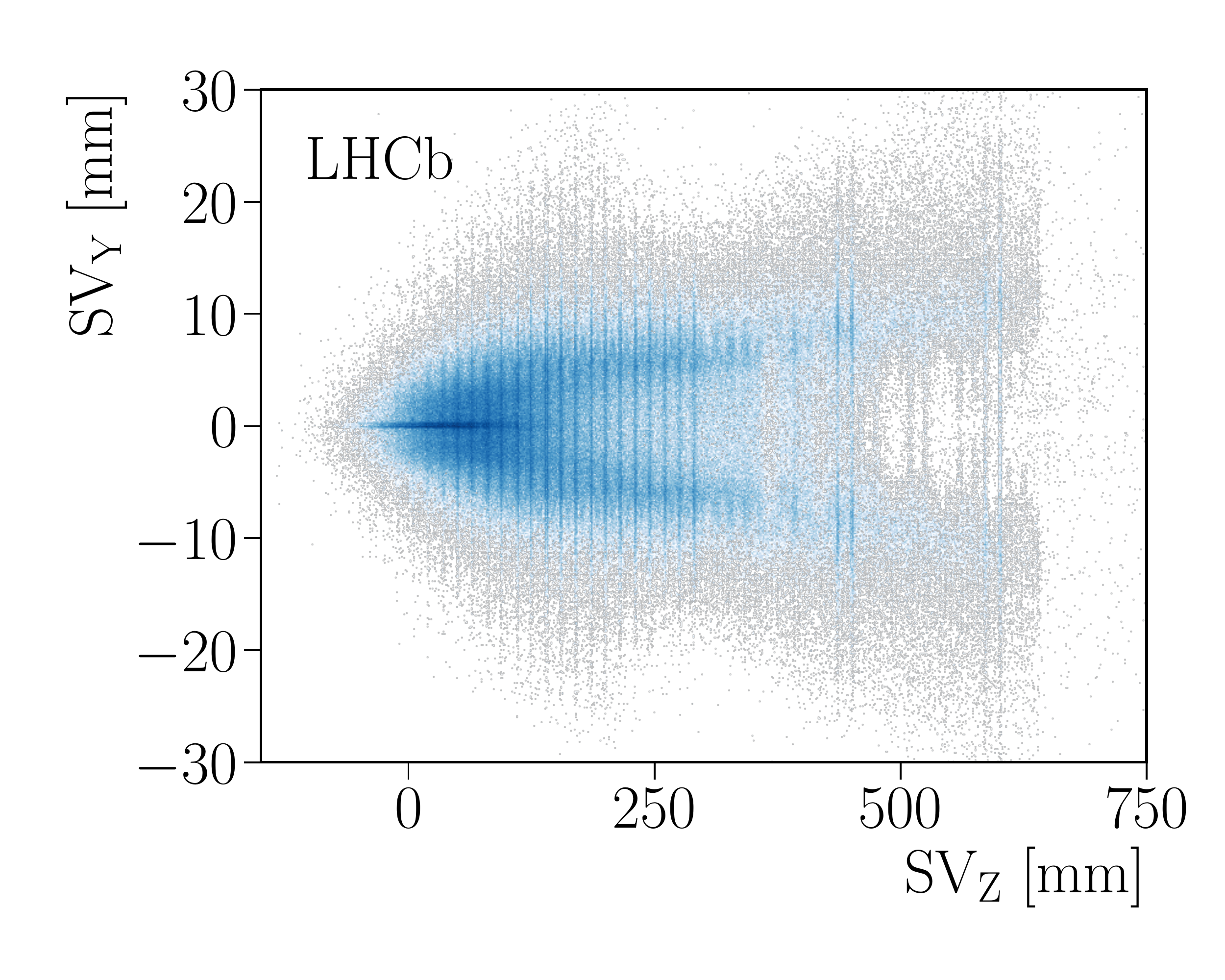}}
  \caption{%
    Position of the secondary vertices for \Ksmumu candidates satisfying the requirement \(m_{\mumu} > 520~\mevcc\).
    The pattern of the innermost subdetector of the \lhcb, the \velo, can be seen, together with that of the surrounding cavity.
  }
  \label{fig:supp:matterveto}
\end{figure}

In the right sideband of the dimuon invariant mass spectrum, candidates originated from material interactions with the detector dominate, as can be seen in Fig.~\ref{fig:supp:matterveto}.
In order to reduce this contribution, a tool profiting from the parametrization of the \velo using proton-gas events, described in detail in Ref.~\cite{LHCb-DP-2018-002}, is used.
This algorithm defines an uncertainty-weighted distance to the material
\begin{equation*}
  D = \sqrt{\left(\frac{x - \text{SV}_x}{\sigma_x}\right)^2 +
    \left(\frac{y - \text{SV}_y}{\sigma_y}\right)^2 +
    \left(\frac{z - \text{SV}_z}{\sigma_z}\right)^2},
\end{equation*}
where \(\text{SV}_{x,y,z}\) denote the position of the reconstructed secondary vertex in the three coordinates, and \(\sigma_{x,y,z}\) the associated uncertainty.
This quantity gives information about how likely a vertex arises from an inelastic material interaction.

\FloatBarrier
\addcontentsline{toc}{section}{References}
\bibliographystyle{LHCb}
\bibliography{main,standard,LHCb-PAPER,LHCb-CONF,LHCb-DP,LHCb-TDR}

\ifx\mcitethebibliography\mciteundefinedmacro
\PackageError{LHCb.bst}{mciteplus.sty has not been loaded}
{This bibstyle requires the use of the mciteplus package.}\fi
\providecommand{\href}[2]{#2}
\begin{mcitethebibliography}{10}
\mciteSetBstSublistMode{n}
\mciteSetBstMaxWidthForm{subitem}{\alph{mcitesubitemcount})}
\mciteSetBstSublistLabelBeginEnd{\mcitemaxwidthsubitemform\space}
{\relax}{\relax}

\bibitem{Ecker:1991ru}
G.~Ecker and A.~Pich, \ifthenelse{\boolean{articletitles}}{\emph{{The
  longitudinal muon polarization in $K_{\rm L}\to \mu^+\mu^-$}},
  }{}\href{https://doi.org/10.1016/0550-3213(91)90056-4}{Nucl.\ Phys.\
  \textbf{B366} (1991) 189}\relax
\mciteBstWouldAddEndPuncttrue
\mciteSetBstMidEndSepPunct{\mcitedefaultmidpunct}
{\mcitedefaultendpunct}{\mcitedefaultseppunct}\relax
\EndOfBibitem
\bibitem{Isidori:2003ts}
G.~Isidori and R.~Unterdorfer, \ifthenelse{\boolean{articletitles}}{\emph{{On
  the short distance constraints from $K_{\rm L,S}\to \mu^+\mu^-$}},
  }{}\href{https://doi.org/10.1088/1126-6708/2004/01/009}{JHEP \textbf{01}
  (2004) 009}, \href{http://arxiv.org/abs/hep-ph/0311084}{{\normalfont\ttfamily
  arXiv:hep-ph/0311084}}\relax
\mciteBstWouldAddEndPuncttrue
\mciteSetBstMidEndSepPunct{\mcitedefaultmidpunct}
{\mcitedefaultendpunct}{\mcitedefaultseppunct}\relax
\EndOfBibitem
\bibitem{DAmbrosio:2017klp}
G.~D'Ambrosio and T.~Kitahara,
  \ifthenelse{\boolean{articletitles}}{\emph{{Direct \CP violation in $K \to
  \mu^+ \mu^-$}},
  }{}\href{https://doi.org/10.1103/PhysRevLett.119.201802}{Phys.\ Rev.\ Lett.\
  \textbf{119} (2017) 201802},
  \href{http://arxiv.org/abs/1707.06999}{{\normalfont\ttfamily
  arXiv:1707.06999}}\relax
\mciteBstWouldAddEndPuncttrue
\mciteSetBstMidEndSepPunct{\mcitedefaultmidpunct}
{\mcitedefaultendpunct}{\mcitedefaultseppunct}\relax
\EndOfBibitem
\bibitem{D'Ambrosio:1994ae}
G.~D'Ambrosio, G.~Ecker, G.~Isidori, and H.~Neufeld,
  \ifthenelse{\boolean{articletitles}}{\emph{{Radiative nonleptonic kaon
  decays}}, }{} in {\em {2nd DAPHNE Physics Handbook}}, 265, 1994,
  \href{http://arxiv.org/abs/hep-ph/9411439}{{\normalfont\ttfamily
  arXiv:hep-ph/9411439}}\relax
\mciteBstWouldAddEndPuncttrue
\mciteSetBstMidEndSepPunct{\mcitedefaultmidpunct}
{\mcitedefaultendpunct}{\mcitedefaultseppunct}\relax
\EndOfBibitem
\bibitem{PDG2018}
Particle Data Group, M.~Tanabashi {\em et~al.},
  \ifthenelse{\boolean{articletitles}}{\emph{{\href{http://pdg.lbl.gov/}{Review
  of particle physics}}},
  }{}\href{https://doi.org/10.1103/PhysRevD.98.030001}{Phys.\ Rev.\
  \textbf{D98} (2018) 030001}\relax
\mciteBstWouldAddEndPuncttrue
\mciteSetBstMidEndSepPunct{\mcitedefaultmidpunct}
{\mcitedefaultendpunct}{\mcitedefaultseppunct}\relax
\EndOfBibitem
\bibitem{Ambrose:2000gj}
E871 collaboration, D.~Ambrose {\em et~al.},
  \ifthenelse{\boolean{articletitles}}{\emph{{Improved branching ratio
  measurement for the decay $\KL\to\mu^{+}\mu^{-}$}},
  }{}\href{https://doi.org/10.1103/PhysRevLett.84.1389}{Phys.\ Rev.\ Lett.\
  \textbf{84} (2000) 1389}\relax
\mciteBstWouldAddEndPuncttrue
\mciteSetBstMidEndSepPunct{\mcitedefaultmidpunct}
{\mcitedefaultendpunct}{\mcitedefaultseppunct}\relax
\EndOfBibitem
\bibitem{Akagi:1994bb}
E137 collaboration, T.~Akagi {\em et~al.},
  \ifthenelse{\boolean{articletitles}}{\emph{{Experimental study of the rare
  decays $\KL \to \mu e$, $\KL \to e e$, $\KL \to \mu\mu$ and $\KL \to e e e
  e$}}, }{}\href{https://doi.org/10.1103/PhysRevD.51.2061}{Phys.\ Rev.\
  \textbf{D51} (1995) 2061}\relax
\mciteBstWouldAddEndPuncttrue
\mciteSetBstMidEndSepPunct{\mcitedefaultmidpunct}
{\mcitedefaultendpunct}{\mcitedefaultseppunct}\relax
\EndOfBibitem
\bibitem{Heinson:1994kq}
E791 collaboration, A.~Heinson {\em et~al.},
  \ifthenelse{\boolean{articletitles}}{\emph{{Measurement of the branching
  ratio for the rare decay $\KL\to\mu^{+}\mu^{-}$}},
  }{}\href{https://doi.org/10.1103/PhysRevD.51.985}{Phys.\ Rev.\  \textbf{D51}
  (1995) 985}\relax
\mciteBstWouldAddEndPuncttrue
\mciteSetBstMidEndSepPunct{\mcitedefaultmidpunct}
{\mcitedefaultendpunct}{\mcitedefaultseppunct}\relax
\EndOfBibitem
\bibitem{Chobanova:2017rkj}
V.~Chobanova {\em et~al.}, \ifthenelse{\boolean{articletitles}}{\emph{{Probing
  SUSY effects in $K_S^0\rightarrow\mu^+\mu^-$}},
  }{}\href{https://doi.org/10.1007/JHEP05(2018)024}{JHEP \textbf{05} (2018)
  024}, \href{http://arxiv.org/abs/1711.11030}{{\normalfont\ttfamily
  arXiv:1711.11030}}\relax
\mciteBstWouldAddEndPuncttrue
\mciteSetBstMidEndSepPunct{\mcitedefaultmidpunct}
{\mcitedefaultendpunct}{\mcitedefaultseppunct}\relax
\EndOfBibitem
\bibitem{Dorsner:2011ai}
I.~Dor\v{s}ner {\em et~al.}, \ifthenelse{\boolean{articletitles}}{\emph{{Limits
  on scalar leptoquark interactions and consequences for GUTs}},
  }{}\href{https://doi.org/10.1007/JHEP11(2011)002}{JHEP \textbf{11} (2011)
  002}, \href{http://arxiv.org/abs/1107.5393}{{\normalfont\ttfamily
  arXiv:1107.5393}}\relax
\mciteBstWouldAddEndPuncttrue
\mciteSetBstMidEndSepPunct{\mcitedefaultmidpunct}
{\mcitedefaultendpunct}{\mcitedefaultseppunct}\relax
\EndOfBibitem
\bibitem{Bobeth:2017ecx}
C.~Bobeth and A.~J. Buras,
  \ifthenelse{\boolean{articletitles}}{\emph{{Leptoquarks meet
  $\varepsilon'/\varepsilon$ and rare kaon processes}},
  }{}\href{https://doi.org/10.1007/JHEP02(2018)101}{JHEP \textbf{02} (2018)
  101}, \href{http://arxiv.org/abs/1712.01295}{{\normalfont\ttfamily
  arXiv:1712.01295}}\relax
\mciteBstWouldAddEndPuncttrue
\mciteSetBstMidEndSepPunct{\mcitedefaultmidpunct}
{\mcitedefaultendpunct}{\mcitedefaultseppunct}\relax
\EndOfBibitem
\bibitem{LHCb-PAPER-2017-009}
LHCb collaboration, R.~Aaij {\em et~al.},
  \ifthenelse{\boolean{articletitles}}{\emph{{Improved limit on the branching
  fraction of the rare decay \mbox{\decay{\KS}{\mup\mun}}}},
  }{}\href{https://doi.org/10.1140/epjc/s10052-017-5230-x}{Eur.\ Phys.\ J.\
  \textbf{C77} (2017) 678},
  \href{http://arxiv.org/abs/1706.00758}{{\normalfont\ttfamily
  arXiv:1706.00758}}\relax
\mciteBstWouldAddEndPuncttrue
\mciteSetBstMidEndSepPunct{\mcitedefaultmidpunct}
{\mcitedefaultendpunct}{\mcitedefaultseppunct}\relax
\EndOfBibitem
\bibitem{Junior:2018odx}
A.~A. Alves~Jr.\ {\em et~al.},
  \ifthenelse{\boolean{articletitles}}{\emph{{Prospects for measurements with
  strange hadrons at LHCb}},
  }{}\href{https://doi.org/10.1007/JHEP05(2019)048}{JHEP \textbf{05} (2019)
  048}, \href{http://arxiv.org/abs/1808.03477}{{\normalfont\ttfamily
  arXiv:1808.03477}}\relax
\mciteBstWouldAddEndPuncttrue
\mciteSetBstMidEndSepPunct{\mcitedefaultmidpunct}
{\mcitedefaultendpunct}{\mcitedefaultseppunct}\relax
\EndOfBibitem
\bibitem{LHCb-DP-2012-002}
A.~A. Alves~Jr.\ {\em et~al.},
  \ifthenelse{\boolean{articletitles}}{\emph{{Performance of the LHCb muon
  system}}, }{}\href{https://doi.org/10.1088/1748-0221/8/02/P02022}{JINST
  \textbf{8} (2013) P02022},
  \href{http://arxiv.org/abs/1211.1346}{{\normalfont\ttfamily
  arXiv:1211.1346}}\relax
\mciteBstWouldAddEndPuncttrue
\mciteSetBstMidEndSepPunct{\mcitedefaultmidpunct}
{\mcitedefaultendpunct}{\mcitedefaultseppunct}\relax
\EndOfBibitem
\bibitem{LHCb-DP-2014-002}
LHCb collaboration, R.~Aaij {\em et~al.},
  \ifthenelse{\boolean{articletitles}}{\emph{{LHCb detector performance}},
  }{}\href{https://doi.org/10.1142/S0217751X15300227}{Int.\ J.\ Mod.\ Phys.\
  \textbf{A30} (2015) 1530022},
  \href{http://arxiv.org/abs/1412.6352}{{\normalfont\ttfamily
  arXiv:1412.6352}}\relax
\mciteBstWouldAddEndPuncttrue
\mciteSetBstMidEndSepPunct{\mcitedefaultmidpunct}
{\mcitedefaultendpunct}{\mcitedefaultseppunct}\relax
\EndOfBibitem
\bibitem{LHCb-DP-2012-004}
R.~Aaij {\em et~al.}, \ifthenelse{\boolean{articletitles}}{\emph{{The \lhcb
  trigger and its performance in 2011}},
  }{}\href{https://doi.org/10.1088/1748-0221/8/04/P04022}{JINST \textbf{8}
  (2013) P04022}, \href{http://arxiv.org/abs/1211.3055}{{\normalfont\ttfamily
  arXiv:1211.3055}}\relax
\mciteBstWouldAddEndPuncttrue
\mciteSetBstMidEndSepPunct{\mcitedefaultmidpunct}
{\mcitedefaultendpunct}{\mcitedefaultseppunct}\relax
\EndOfBibitem
\bibitem{Dettori:2297352}
F.~Dettori, D.~Mart\'{\i}nez~Santos, and J.~Prisciandaro,
  \ifthenelse{\boolean{articletitles}}{\emph{{Low-\pt dimuon triggers at LHCb
  in Run 2}}, }{}
  \href{http://cdsweb.cern.ch/search?p=LHCb-PUB-2017-023&f=reportnumber&action_search=Search&c=LHCb+Notes}
  {LHCb-PUB-2017-023}, 2017\relax
\mciteBstWouldAddEndPuncttrue
\mciteSetBstMidEndSepPunct{\mcitedefaultmidpunct}
{\mcitedefaultendpunct}{\mcitedefaultseppunct}\relax
\EndOfBibitem
\bibitem{LHCb-TDR-018}
LHCb collaboration, \ifthenelse{\boolean{articletitles}}{\emph{{Computing Model
  of the Upgrade LHCb experiment}}, }{}
  \href{http://cdsweb.cern.ch/search?p=CERN-LHCC-2018-014&f=reportnumber&action_search=Search&c=LHCb+Reports}
  {CERN-LHCC-2018-014}, 2018\relax
\mciteBstWouldAddEndPuncttrue
\mciteSetBstMidEndSepPunct{\mcitedefaultmidpunct}
{\mcitedefaultendpunct}{\mcitedefaultseppunct}\relax
\EndOfBibitem
\bibitem{LHCb-2008-073}
J.~A. Hernando~Morata {\em et~al.},
  \ifthenelse{\boolean{articletitles}}{\emph{{Measurement of trigger
  efficiencies and biases}}, }{}
  \href{http://cdsweb.cern.ch/search?p=LHCb-2008-073&f=reportnumber&action_search=Search&c=LHCb+Notes}
  {LHCb-2008-073}, 2010\relax
\mciteBstWouldAddEndPuncttrue
\mciteSetBstMidEndSepPunct{\mcitedefaultmidpunct}
{\mcitedefaultendpunct}{\mcitedefaultseppunct}\relax
\EndOfBibitem
\bibitem{armenteros}
J.~Podolanski and R.~Armenteros,
  \ifthenelse{\boolean{articletitles}}{\emph{{Analysis of V-events}},
  }{}\href{https://doi.org/10.1080/14786440108520416}{Phil.\ Mag.\  \textbf{45}
  (1954) 13}\relax
\mciteBstWouldAddEndPuncttrue
\mciteSetBstMidEndSepPunct{\mcitedefaultmidpunct}
{\mcitedefaultendpunct}{\mcitedefaultseppunct}\relax
\EndOfBibitem
\bibitem{LHCb-DP-2013-001}
F.~Archilli {\em et~al.},
  \ifthenelse{\boolean{articletitles}}{\emph{{Performance of the muon
  identification at LHCb}},
  }{}\href{https://doi.org/10.1088/1748-0221/8/10/P10020}{JINST \textbf{8}
  (2013) P10020}, \href{http://arxiv.org/abs/1306.0249}{{\normalfont\ttfamily
  arXiv:1306.0249}}\relax
\mciteBstWouldAddEndPuncttrue
\mciteSetBstMidEndSepPunct{\mcitedefaultmidpunct}
{\mcitedefaultendpunct}{\mcitedefaultseppunct}\relax
\EndOfBibitem
\bibitem{AdaBoost}
Y.~Freund and R.~E. Schapire, \ifthenelse{\boolean{articletitles}}{\emph{A
  decision-theoretic generalization of on-line learning and an application to
  boosting}, }{}\href{https://doi.org/10.1006/jcss.1997.1504}{J.\ Comput.\
  Syst.\ Sci.\  \textbf{55} (1997) 119}\relax
\mciteBstWouldAddEndPuncttrue
\mciteSetBstMidEndSepPunct{\mcitedefaultmidpunct}
{\mcitedefaultendpunct}{\mcitedefaultseppunct}\relax
\EndOfBibitem
\bibitem{Breiman}
L.~Breiman, J.~H. Friedman, R.~A. Olshen, and C.~J. Stone, {\em Classification
  and regression trees}, Wadsworth international group, Belmont, California,
  USA, 1984\relax
\mciteBstWouldAddEndPuncttrue
\mciteSetBstMidEndSepPunct{\mcitedefaultmidpunct}
{\mcitedefaultendpunct}{\mcitedefaultseppunct}\relax
\EndOfBibitem
\bibitem{XGBoost}
T.~Chen and C.~Guestrin, \ifthenelse{\boolean{articletitles}}{\emph{{XGB}oost:
  A scalable tree boosting system}, }{} in {\em Proceedings of the 22Nd ACM
  SIGKDD International Conference on Knowledge Discovery and Data Mining},
  \href{https://doi.org/10.1145/2939672.2939785}{ KDD '16, (New York, NY, USA),
  785--794, ACM, 2016},
  \href{http://arxiv.org/abs/1603.02754}{{\normalfont\ttfamily
  arXiv:1603.02754}}\relax
\mciteBstWouldAddEndPuncttrue
\mciteSetBstMidEndSepPunct{\mcitedefaultmidpunct}
{\mcitedefaultendpunct}{\mcitedefaultseppunct}\relax
\EndOfBibitem
\bibitem{Rogozhnikov:2016bdp}
A.~Rogozhnikov, \ifthenelse{\boolean{articletitles}}{\emph{{Reweighting with
  Boosted Decision Trees}},
  }{}\href{https://doi.org/10.1088/1742-6596/762/1/012036}{J.\ Phys.\ Conf.\
  Ser.\  \textbf{762} (2016) },
  \href{http://arxiv.org/abs/1608.05806}{{\normalfont\ttfamily
  arXiv:1608.05806}}, \url{https://github.com/arogozhnikov/hep_ml}\relax
\mciteBstWouldAddEndPuncttrue
\mciteSetBstMidEndSepPunct{\mcitedefaultmidpunct}
{\mcitedefaultendpunct}{\mcitedefaultseppunct}\relax
\EndOfBibitem
\bibitem{scikit-learn}
F.~Pedregosa {\em et~al.},
  \ifthenelse{\boolean{articletitles}}{\emph{Scikit-learn: Machine learning in
  {P}ython}, }{}\href{http://www.jmlr.org/papers/v12/pedregosa11a.html}{Journal
  of Machine Learning Research \textbf{12} (2011) 2825}, see also
  \url{http://scikit-learn.org}\relax
\mciteBstWouldAddEndPuncttrue
\mciteSetBstMidEndSepPunct{\mcitedefaultmidpunct}
{\mcitedefaultendpunct}{\mcitedefaultseppunct}\relax
\EndOfBibitem
\bibitem{LHCb-DP-2018-002}
M.~Alexander {\em et~al.}, \ifthenelse{\boolean{articletitles}}{\emph{{Mapping
  the material in the LHCb vertex locator using secondary hadronic
  interactions}},
  }{}\href{https://doi.org/10.1088/1748-0221/13/06/P06008}{JINST \textbf{13}
  (2018) P06008}, \href{http://arxiv.org/abs/1803.07466}{{\normalfont\ttfamily
  arXiv:1803.07466}}\relax
\mciteBstWouldAddEndPuncttrue
\mciteSetBstMidEndSepPunct{\mcitedefaultmidpunct}
{\mcitedefaultendpunct}{\mcitedefaultseppunct}\relax
\EndOfBibitem
\bibitem{Babusci:2019gqu}
KLOE-2 collaboration, D.~Babusci {\em et~al.},
  \ifthenelse{\boolean{articletitles}}{\emph{{Measurement of the branching
  fraction for the decay $K_S \to \pi \mu \nu$ with the KLOE detector}},
  }{}\href{http://arxiv.org/abs/1912.05990}{{\normalfont\ttfamily
  arXiv:1912.05990}}\relax
\mciteBstWouldAddEndPuncttrue
\mciteSetBstMidEndSepPunct{\mcitedefaultmidpunct}
{\mcitedefaultendpunct}{\mcitedefaultseppunct}\relax
\EndOfBibitem
\bibitem{Hypatia}
{D.\ ~Mart\'inez Santos and F.\ ~Dupertuis},
  \ifthenelse{\boolean{articletitles}}{\emph{{Mass distributions marginalized
  over per-event errors}},
  }{}\href{https://doi.org/10.1016/j.nima.2014.06.081}{Nucl.\ Instrum.\ Meth.\
  \textbf{A764} (2014) 150},
  \href{http://arxiv.org/abs/1312.5000}{{\normalfont\ttfamily
  arXiv:1312.5000}}\relax
\mciteBstWouldAddEndPuncttrue
\mciteSetBstMidEndSepPunct{\mcitedefaultmidpunct}
{\mcitedefaultendpunct}{\mcitedefaultseppunct}\relax
\EndOfBibitem
\bibitem{LHCb-PAPER-2012-023}
LHCb collaboration, R.~Aaij {\em et~al.},
  \ifthenelse{\boolean{articletitles}}{\emph{{Search for the rare decay
  \mbox{\decay{\KS}{\mumu}}}},
  }{}\href{https://doi.org/10.1007/JHEP01(2013)090}{JHEP \textbf{01} (2013)
  090}, \href{http://arxiv.org/abs/1209.4029}{{\normalfont\ttfamily
  arXiv:1209.4029}}\relax
\mciteBstWouldAddEndPuncttrue
\mciteSetBstMidEndSepPunct{\mcitedefaultmidpunct}
{\mcitedefaultendpunct}{\mcitedefaultseppunct}\relax
\EndOfBibitem
\end{mcitethebibliography}

\newpage
\centerline
{\large\bf LHCb collaboration}
\begin
{flushleft}
\small
R.~Aaij$^{31}$,
C.~Abell{\'a}n~Beteta$^{49}$,
T.~Ackernley$^{59}$,
B.~Adeva$^{45}$,
M.~Adinolfi$^{53}$,
H.~Afsharnia$^{9}$,
C.A.~Aidala$^{80}$,
S.~Aiola$^{25}$,
Z.~Ajaltouni$^{9}$,
S.~Akar$^{66}$
% ...

{\footnotesize \it

% ...
$ ^{9}$Universit{\'e} Clermont Auvergne, CNRS/IN2P3, LPC, Clermont-Ferrand, France\\
% ...
$ ^{25}$INFN Sezione di Milano, Milano, Italy\\
% ...
$ ^{31}$Nikhef National Institute for Subatomic Physics, Amsterdam, Netherlands\\
% ...
$ ^{45}$Instituto Galego de F{\'\i}sica de Altas Enerx{\'\i}as (IGFAE), Universidade de Santiago de Compostela, Santiago de Compostela, Spain\\
% ...
$ ^{49}$Physik-Institut, Universit{\"a}t Z{\"u}rich, Z{\"u}rich, Switzerland\\
% ...
$ ^{53}$H.H. Wills Physics Laboratory, University of Bristol, Bristol, United Kingdom\\
% ...
$ ^{59}$Oliver Lodge Laboratory, University of Liverpool, Liverpool, United Kingdom\\
% ...
$ ^{66}$Los Alamos National Laboratory (LANL), Los Alamos, United States\\
$ ^{67}$Syracuse University, Syracuse, NY, United States\\
% ...
$ ^{80}$University of Michigan, Ann Arbor, United States, associated to $^{67}$\\
\bigskip
% ...
\medskip
}
\end{flushleft}

\end{document}